\documentclass[twocolumn]{elsart}
\newcommand{\arcsecs}{\mbox{$^{\prime\prime}$}}
\usepackage{natbib}
\usepackage{graphicx}
\usepackage{amsmath,amssymb}
\journal{New Astronomy Reviews}
\begin{document}
\runauthor{Jakobsson \& Fynbo}
\begin{frontmatter}
\title{What made GRBs 060505 and 060614?}
\author[Herts]{P\'all Jakobsson},
\author[Copenhagen]{Johan P. U. Fynbo}

\address[Herts]{Centre for Astrophysics Research, University of Hertfordshire,
College Lane, Hatfield, Herts, AL10 9AB, UK}
\address[Copenhagen]{Dark Cosmology Centre, Niels Bohr Institute, University 
of Copenhagen, Juliane Maries Vej 30, 2100 Copenhagen, Denmark}

\begin{abstract}
Recent observations of two nearby SN-less long-duration gamma-ray bursts 
(GRBs), which share no obvious characteristics in their prompt emission, 
suggest a new phenomenological type of massive stellar death. Here we briefly 
review the observational properties of these bursts and their proposed hosts, 
and discuss whether a new GRB classification scheme is needed.
\end{abstract}
\end{frontmatter}

\section{Introduction}
\begin{figure*}
\centering
\includegraphics[width=0.99\textwidth]{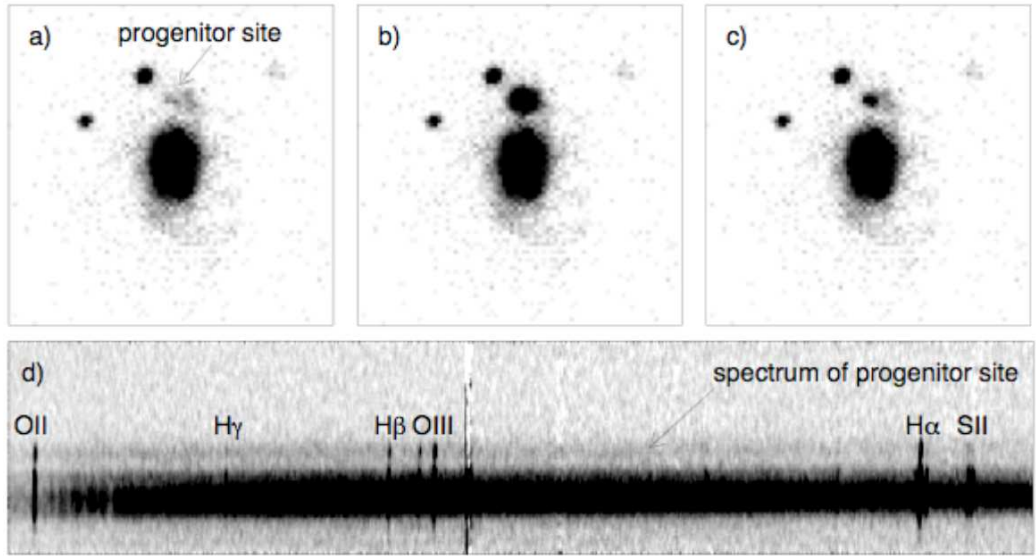}
\caption{(a) The field (20\arcsecs $\times$ 20\arcsecs) of GRB\,060505 as
observed from the VLT in the $R$-band on 22 May 2006. The arrow marks the 
position where the optical afterglow was detected in earlier imaging. 
(b) As the image would have looked had a SN like 1998bw been present in the 
data. The strict upper limits strongly exclude the bright SNe 1998bw and 
2006aj that were associated with long GRBs. (c) Similar to (b), but with a 
very faint Ic SN, such as 2002ap, added. (d) The 2-D optical spectrum 
obtained with VLT/FORS2. The slit covered the centre of the host galaxy 
and the location of GRB\,060505. As seen in the spectrum, this site is a 
bright star-forming region in the host galaxy suggesting that the progenitor 
was a massive star.}
\label{060505.fig}
\end{figure*}

A broad-lined and luminous type Ic core-collapse supernova (SN) is predicted
to accompany every long-duration gamma-ray burst (GRB) in the standard 
collapsar model \citep{woosley}. Although this association had been confirmed
in observations of several nearby GRBs \citep[e.g.][]{jens}, a new controversy
commenced when no SN emission accompanied GRBs 060505 ($z = 0.09$, 
duration $\sim$4\,s) and 060614 ($z = 0.13$, duration $\sim$100\,s) down to
limits fainter than any known type Ic SN and hundreds of times fainter than
the archetypal SN\,1998bw \citep{della,johan,gal-yam}. The upper panels of
Fig.\ref{060505.fig} illustrate how easily such SNe would have been detected
in the case of GRB\,060505.
\par
An important clue to the origin and progenitors of these bursts, is the
nature of the host galaxies. The GRB\,060505 host is a spiral galaxy, 
atypical for long-duration bursts but not unheard of (GRB\,980425: 
Fynbo et al., 2000; GRB\,990705: Le Floc'h et al., 2002; GRB\,020819: 
Jakobsson et al., 2005). The burst occurred inside a compact star-forming 
\mbox{H\,{\scriptsize II}} region in one of the spiral arms, and a spatially 
resolved spectroscopy (lower panel of Fig.\ref{060505.fig}) revealed that 
the properties of the GRB site are similar to those found for other 
long-duration GRBs with a high specific star formation rate (SSFR) and low 
metallicity \citep{CT}. The GRB\,060614 host is significantly fainter 
(one of the least luminous GRB host ever detected) with a moderate SSFR.
\section{Discussion}
\subsection{High extinction?}
Could the emission from an associated SN be completely obscured by dust along
the line-of-sight? The levels of Galactic extinction are very low in both
directions. Host extinction of more than a magnitude is also unlikely in
either case since the host galaxy spectra display no reddening as derived
from the Balmer line ratios. In addition, the GRB\,060614 afterglow is
clearly detected in the UV \citep{holland}.
\subsection{Wrong redshifts?}
Another option is that the proposed host galaxies are chance encounters
along the line-of-sight \citep{cobb,schaefer}, and the real GRB redshifts
are much higher (rendering a SN too faint to be observed). However, a few
observational facts argue against this scenario. In the case of GRB\,060614:
(i) the UV detection places an upper limit of around 1.1 on the redshift;
(ii) no absorption components in the optical afterglow spectrum 
\citep{fugazza}, as expected for a low redshift, but not for a high-$z$
burst with a foreground galaxy; (iii) very deep HST images of the field
should have revealed the ``true host'' at $z \lesssim 1.1$, but none was
seen \citep{gal-yam}. For GRB\,060505 it is extremely unlikely that the
afterglow accidentally superposed right on top of a small star-forming
region within a foreground spiral galaxy.
\subsection{No SNe: a problem?}
The host galaxies and the GRB location within them strongly suggest
an association with star formation, and hence a massive stellar origin. It
is important to realize that the lack of a strong SN emission was actually 
predicted as a variant of the original collapsar model, e.g. collapse of a 
massive star with an explosion energy so small that most of the $^{56}$Ni 
falls back into the black hole \citep[e.g.][]{heger,fryer}. In
another variant of the collapsar model, progenitor stars with relatively
low angular momentum could also produce SN-less GRBs \citep{macfadyen}.
\par
We should also remember that the duration distributions of short and long
GRBs overlap. In fact, the GRB\,060505 duration of 4\,s is near the 
$\sim$5\,s duration which \citet{donaghy} find as the point of roughly
equal probability of a given burst lying in either the short or long class.
It has been suggested that the physical mechanism for this burst is the
same as for short bursts, i.e. a merger of compact objects \citep{ofek},
although the progenitor time delay of only $\lesssim 7$\,Myr is on the 
borderline for allowed values \citep{CT}. However, such short time delays
have been proposed via newly recognized formation channels, which lead to
the formation of tighter double compact objects with short lifetimes and
therefore possible prompt merger within hosts \citep{bel3}. Whether such
channels require a low metallicity as found for GRB\,060505 \citep{CT}
remains to be explored.
\subsection{Classification problem?}
With the added complication that the $\sim$100\,s long GRB\,060614 is
located among the short bursts in the lag-luminosity plot, it has been 
argued that a new GRB classification scheme is required \citep{gehrels}.
We do not think this is the case, as the current GRB classification is
operationally well defined. Rather that new observations are warning us not
necessarily to expect a very simple mapping between the duration of the GRB
and the nature of the progenitor: long bursts ($>$2\,s) synonymous with 
massive stars and short bursts ($<$2\,s) synonymous with compact object
mergers.
\par
Others want to abandon the long-short paradigm altogether due to these
``oddball'' bursts, and invent a new terminology: Type I and II bursts
similar to the SN classification scheme \citep{zhang}. In this scheme,
eight different properties have to be considered for each burst/host.
However, this scheme can be ambiguous (e.g. GRB\,060505) and is not
operational, i.e. involves observables that are not available for most
bursts (associated SN). Using proposed hosts (i.e. a nearby bright galaxy) 
to make a distinction between the two burst populations can also be risky
\citep[e.g. GRB\,060912A: ][]{levan}. In addition, one might envisage a 
Type III category consisting of the new type of bursts (massive white 
dwarf/neutron star merger) suggested by \citet{king}. These could produce 
long bursts definitely without an accompanying SN and have a strong 
correlation with star formation. However, rare members of the class need 
not be near star-forming regions, and could have any type of host galaxy.
\par
It is clear that the two SN-less long bursts from last summer have
raised a few warning flags, i.e. how we think about the long/short
dichotomy. At this point in time, we only recommend that people keep an 
open mind.

\ack
We thank all the co-authors of the \cite{johan} paper. PJ acknowledges 
support by a Marie Curie Intra-European Fellowship within the 6th European 
Community Framework Program under contract number MEIF-CT-2006-042001.

\end{document}